\documentclass[%
aip,
apl,%
amsmath,amssymb,
reprint,superscriptaddress,footinbib,%
]{revtex4-1}

\usepackage{graphicx}
\usepackage{bm}

\usepackage{siunitx}

\begin{document}
	
	
	\title{Spin Hall magnetoresistance in heterostructures consisting of noncrystalline paramagnetic YIG and Pt}
	
	\author{Michaela Lammel}
	\email{m.lammel@ifw-dresden.de}
	\affiliation{Institute for Metallic Materials, Leibnitz Institute of Solid State and Materials Science, 01069 Dresden, Germany}
	\affiliation{Technische Universit\"at Dresden, Institute of Applied Physics, 01062 Dresden, Germany}
	
	\author{Richard Schlitz}
	\affiliation{Institut f\"ur Festk\"orper- und Materialphysik, Technische Universit\"at Dresden, 01062 Dresden, Germany}
	
	\author{Kevin Geishendorf}
	\affiliation{Institute for Metallic Materials, Leibnitz Institute of Solid State and Materials Science, 01069 Dresden, Germany}
	\affiliation{Technische Universit\"at Dresden, Institute of Applied Physics, 01062 Dresden, Germany}
	
	\author{Denys Makarov}
	\affiliation{Helmholtz-Zentrum Dresden-Rossendorf e.V., Institute of Ion Beam Physics and Materials Research, 01328 Dresden, Germany}%
	
	\author{Tobias Kosub}
	\affiliation{Helmholtz-Zentrum Dresden-Rossendorf e.V., Institute of Ion Beam Physics and Materials Research, 01328 Dresden, Germany}%
	
	\author{Savio Fabretti}
	\affiliation{Institut f\"ur Festk\"orper- und Materialphysik, Technische Universit\"at Dresden, 01062 Dresden, Germany}
	
	\author{Helena Reichlova}
	\affiliation{Institut f\"ur Festk\"orper- und Materialphysik, Technische Universit\"at Dresden, 01062 Dresden, Germany}
	
	\author{Rene Huebner}
	\affiliation{Helmholtz-Zentrum Dresden-Rossendorf e.V., Institute of Ion Beam Physics and Materials Research, 01328 Dresden, Germany}%
	
	\author{Kornelius Nielsch}
	\affiliation{Institute for Metallic Materials, Leibnitz Institute of Solid State and Materials Science, 01069 Dresden, Germany}
	\affiliation{Technische Universit\"at Dresden, Institute of Applied Physics, 01062 Dresden, Germany}
	\affiliation{Technische Universit\"at Dresden, Institute of Materials Science, 01062 Dresden, Germany}%
	
	\author{Andy Thomas}
	\affiliation{Institute for Metallic Materials, Leibnitz Institute of Solid State and Materials Science, 01069 Dresden, Germany}
	
	\author{Sebastian T.B.\ Goennenwein}
	\email{sebastian.goennenwein@tu-dresden.de}
	\affiliation{Institut f\"ur Festk\"orper- und Materialphysik, Technische Universit\"at Dresden, 01062 Dresden, Germany}
	
	\date{\today}
	
	\begin{abstract}
		The spin Hall magnetoresistance (SMR) effect arises from spin-transfer processes across the interface between a spin Hall active metal and an insulating magnet. While the SMR response of ferrimagnetic and antiferromagnetic insulators has been studied extensively, the SMR of a paramagnetic spin ensemble is not well established. Thus, we investigate herein the magnetoresistive response of as-deposited yttrium iron garnet/platinum thin film bilayers as a function of the orientation and the amplitude of an externally applied magnetic field. Structural and magnetic characterization show no evidence for crystalline order or spontaneous magnetization in the yttrium iron garnet layer. Nevertheless, we observe a clear magnetoresistance response with a dependence on the magnetic field orientation characteristic for the SMR. We propose two models for the origin of the SMR response in paramagnetic insulator/Pt heterostructures. The first model describes the SMR of an ensemble of non-interacting paramagnetic moments, while the second model describes the magnetoresistance arising by considering the total net moment. Interestingly, our experimental data are consistently described by the net moment picture, in contrast to the situation in compensated ferrimagnets or antiferromagnets.
	\end{abstract}
	\maketitle
	Spin Hall magnetoresistance (SMR)\cite{Chen_Theory_of_SMR,AlthammerSMR,Nakayama_Proximity} is commonly observed in ferrimagnetic insulator (FMI)/normal metal (NM) heterostructures when the metal exhibits a large spin-orbit coupling. The SMR arises due to the interplay of the spin-transfer torque, the spin Hall effect (SHE) and the inverse spin Hall effect at the FMI/NM interface. \cite{Ralph_STT,DYAKONOV_SpinHallEffect,Hirsch_spin_Hall_effect} While the SMR effect is usually discussed in terms of the total (net) magnetization, \cite{Chen_Theory_of_SMR} recent experimental work showed that the SMR does not only probe the net magnetization of FMIs, but is also sensitive to the contributions of the different magnetic sublattices. \cite{Ganzhorn_GdIG,Dong_GdIG} This observation is key to understand the SMR response of more complex magnetic systems, such as canted ferrimagnets \cite{Gepraegs_proximity,Ganzhorn_GdIG,Dong_GdIG}, antiferromagnets \cite{Fischer_NiO,Hoogeboom_NiO,Gomonay_AFM_spintronics,Dazhi_SMRinNiOpusYIG,Han_SMRinAFM_SrMnO3,Ji_Cr2O3plusW}, spin spirals \cite{Aqueel_SMR_paramagnetic-Phase} or helical phases. \cite{Aqeel_helicalSpirals} To date, SMR measurements have been performed extensively in samples with different long-range (spontaneous) magnetic ordering. \cite{AlthammerSMR,Ganzhorn_GdIG,Fischer_NiO,Aqueel_SMR_paramagnetic-Phase,Aqeel_helicalSpirals,Schlitz_Cr2O3} In contrast, paramagnetic materials have not been in the focus of prior work done for SMR measurements. However, the magnetoresistive response of paramagnetic materials is an interesting topic. For example, magnetoresistance measurements were recently performed in a gated paramagnetic ionic liquid. \cite{Liang_paramagnetic-MR} The presence of SMR has been reported by two groups in different magnetically ordered materials, in the paramagnetic phase above the ordering temperature. \cite{Schlitz_Cr2O3,Aqueel_SMR_paramagnetic-Phase} Since the SMR is primarily studied in the magnetically ordered phase in those works, the authors do not provide a microscopic picture for the SMR in a randomly ordered spin ensemble. Therefore, in this work, we systematically study the SMR in a paramagnetic insulator (PMI)/spin Hall metal bilayer and critically compare the experimental results to the SMR expected from two different microscopic models: one model assumes an ensemble of noninteracting moments, while the other model considers the (induced) net magnetization. More specifically, we investigate bilayers fabricated by sputtering of Y$_3$Fe$_5$O$_{12}$ (YIG) and Pt at room temperature. These heterostructures do not show a crystalline order of the YIG layer or spontaneous magnetization, such that we take the YIG layer to be paramagnetic, but they nevertheless exhibit a clear SMR-like magnetoresistive response.
	
\begin{figure}
		\includegraphics[width=\linewidth]{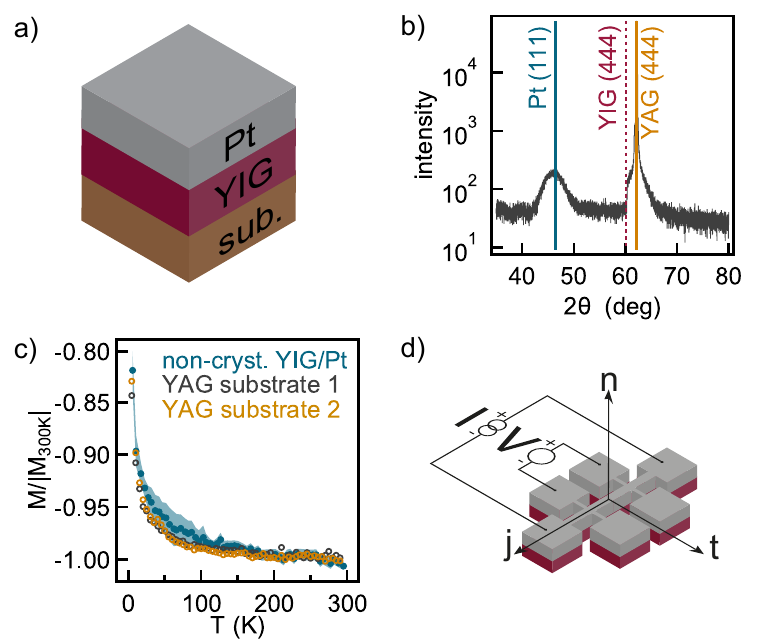}
		\caption{a) Schematic of the YIG/Pt bilayer sample. b) X-ray diffraction $\theta$-$2\theta$ scan of a typical noncrystalline YIG/Pt bilayer. The colored vertical lines give the expected positions for different crystalline diffraction peaks. c) Normalized magnetization of a noncrystalline YIG/Pt bilayer and two YAG substrates as a function of temperature. The light blue shading around the data indicates the scatter of two subsequent measurement runs. The observed moment is negative due to the diamagnetic substrate. d) Electrical contacting scheme of the patterned sample.}
		\label{fig:XRD}
	\end{figure}

	The YIG/Pt bilayers were fabricated via sputtering at room temperature from 2 inch YIG and Pt targets on commercially available (111)-oriented single-crystalline yttrium aluminum garnet (Y$_3$Al$_5$O$_{12}$, YAG) substrates.\cite{AlthammerSMR, Gepraegs_proximity, Ganzhorn_GdIG} To rule out crystallization of the deposited YIG layer on the YAG substrate due to the low lattice mismatch, reference samples were fabricated in the same manner on (100)-oriented Si wafers terminated by a thermal oxide layer of $\SI{1}{\micro\meter}$. The substrates were immersed in isopropanol and ethanol and cleaned in an ultrasonic bath prior to the deposition. The YIG layer was deposited via  RF sputtering at $\SI{80}{\watt}$ for $\SI{6000}{\second}$. Subsequently Pt was deposited using DC sputtering for $\SI{73}{\second}$ at $\SI{30}{\watt}$ without breaking the vacuum. A schematic of a typical stack is given in Fig.\ref{fig:XRD}a. The above-mentioned sputtering parameters resulted in layer thicknesses of $\mathrm{d_{YIG}}=(30\pm1)\,\SI{}{\nano\meter}$ for the YIG layer and $\mathrm{d_{Pt}}= (2.5\pm 0.5)\,$nm for the Pt layer, as confirmed by X-ray reflectometry.
	
	X-ray diffraction measurements were performed using a Bruker D8 Advanced diffractometer equipped with a cobalt anode. As shown in Fig.\ref{fig:XRD}b, we do not observe diffraction peaks that could be linked to YIG. We take this as evidence that the YIG does not grow as small crystallites, but rather as an unordered ``amorphous" layer. Therefore such YIG layers will be referred to as ``noncrystalline" in the following. In contrast, the Pt layer is textured in $\langle111\rangle$-direction which has been reported to be the preferred orientation direction of Pt deposited at room temperature. \cite{Narayan_texturedPt} The sharp step at $2\theta = 60\,$deg results from the iron filter that is used to suppress the Co$_{\mathrm{K}_\beta}$ radiation. TEM studies on the YIG/Pt bilayers additionally confirm the noncrystallinity of the YIG layer while energy-dispersive X-ray spectroscopy analyses show the YIG layer to be stoichometrically identical to the YAG substrate.
	For the magnetic characterization, a Quantum Design MPMS-XL7 SQUID magnetometer with reciprocating sample option was used. Figure \ref{fig:XRD}c shows the magnetization as a function of temperature measured at $\SI{500}{\milli\tesla}$ after cooling the sample in zero magnetic field. As a reference, two substrates were measured by the identical procedure. To ensure comparability and to account for differences in sample size, the data were normalized to the magnetization at \SI{300}{\kelvin}. Comparing the normalized M(T) data from the YIG/Pt bilayer to the bare YAG substrates, we conclude that within the measurement error, no (spontaneous) magnetization of the film can be detected in our samples. Moreover, the samples exhibit only the negative magnetization expected for a diamagnetic substrate. The low temperature paramagnetic-like behavior is likely caused by paramagnetic dopants that were consistently observed in the commercial YAG substrates.\footnote{The measured moment of the substrates corresponds to a concentration of paramagnetic dopants of $\approx10^{-17}\,\mathrm{cm}^{-1}$. Besides, one would expect a magnetic moment of $\approx10^{-8}\,\mathrm{Am}^2$ at $\SI{10}{\kelvin}$ assuming that all Fe moments of YIG are independently contributing whereas a magnetic moment of $\approx10^{-14}\,\mathrm{Am}^2$ is expected for the YAG substrate with assuming a susceptibilty of YAG of $\chi_{\mathrm{YAG}}=10^{-5}$.}
	
	For magnetotransport measurements, Hall bars with a contact separation of $l=\SI{400}{\micro\meter}$ along the direction of current flow and a width of $w=\SI{50}{\micro\meter}$ were defined by using optical lithography and consecutive Ar ion etching. Subsequently, the samples were mounted into a chip carrier and contacted by aluminum wire bonding. The electric contacting scheme as well as the used coordinate system is given in Fig.\ref{fig:XRD}d. A current $I=\SI{90}{\micro\ampere}$ was applied along the Hall bar (along \textbf{j} direction) utilizing a Keithley 2450 sourcemeter. To decrease the noise level and to enhance the sensitivity, a current reversal technique was used. \cite{GoennenweinMMR} The longitudinal voltage $V$, i.e. the voltage drop along the direction of current flow, was measured by a Keithley 2182 nanovoltmeter.
	
\begin{figure}
		\includegraphics[width=\linewidth]{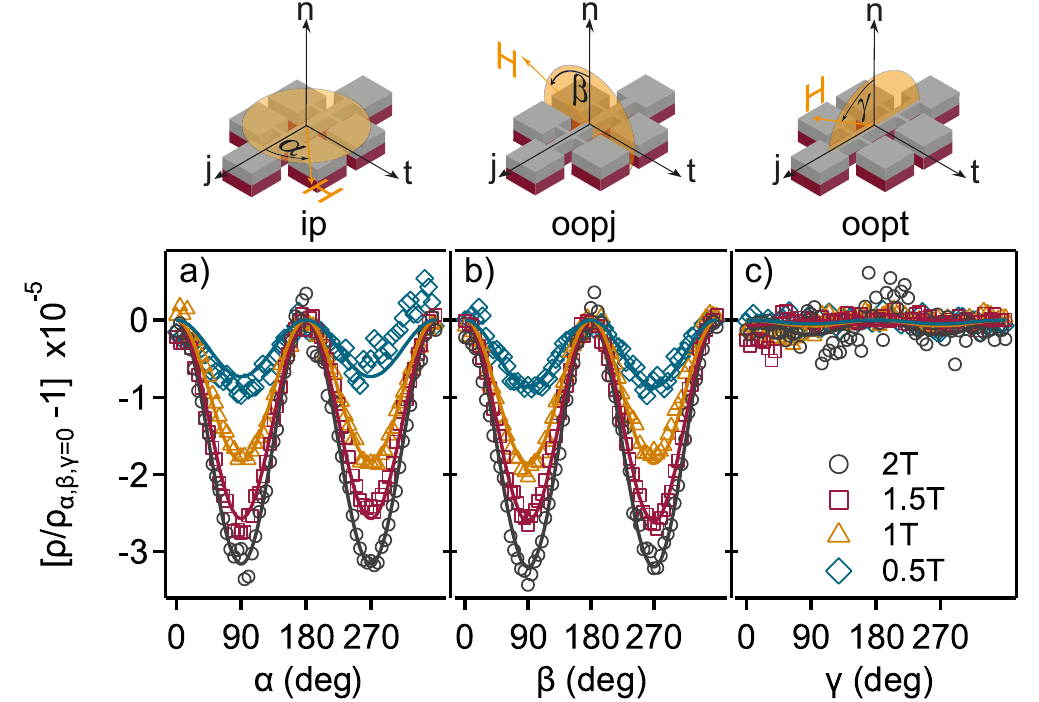}
		\caption{Magnetoresistance measurements as a function of magnetic field orientation, recorded at $200\,$K in three orthogonal rotation planes. The rotation geometries are displayed above the corresponding panel. Measurements were performed at different, fixed magnetic field strengths $\mu_0H=0.5\,$T, $1\,$T, $1.5\,$T and $2\,$T, which are given by the open blue rhombi, yellow triangles, red squares and gray circles, respectively. A $\sin^2(\alpha, \beta, \gamma)$ fit to the data is given by the solid line in the associated color (cf. Eq.\ref{eq:SMR}).}
		\label{fig:ADMR}
	\end{figure}

	Field orientation dependent magnetoresistance measurements at different temperatures and in three orthogonal rotation planes were performed in a 3D vector magnetic field cryostat. The in-plane rotation of a constant external magnetic field \textbf{H} around the surface normal \textbf{n} is herein denoted as ip (angle $\alpha$), the out-of-plane rotation around the current direction \textbf{j} as oopj (angle $\beta$) and the out-of-plane rotation around the \textbf{t} direction as oopt (angle $\gamma$), as is shown above Fig.\ref{fig:ADMR}a, b and c, respectively. In a model FMI/NM system with one single magnetic sublattice pointing along the magnetization unit vector $\mathbf{m}=\frac{\mathbf{M}}{\mathrm{M}}$, the SMR can be described by: \cite{Chen_Theory_of_SMR} 
	\begin{equation}
	\begin{split}
	\rho&=\rho_0 + \Delta\rho\,(1-\mathrm{m_t}^2)\\
	&=\rho_0 + \Delta\rho\,[1-\sin^2(\alpha, \beta)]
	\end{split}
	\label{eq:SMR}
	\end{equation}
	where $\Delta\rho>0$ gives the change of resistance as a function of the projection of the magnetization unit vector $\mathbf{m}$ on the \textbf{t} direction $\mathrm{m_t}$, as defined above. Thus, following Eq.\ref{eq:SMR}, the resistance in a typical SMR measurement is minimal for \textbf{m}$||$\textbf{t} and maximal for \textbf{m}$\perp$\textbf{t}.
	Figure \ref{fig:ADMR} shows the dependence of the magnetoresistance on the angles $\alpha$, $\beta$ and $\gamma$, at \SI{200}{\kelvin} for different amplitudes of the external magnetic field. For each field amplitude, the voltage $V$ is recorded for clockwise and anticlockwise rotation of the magnetic field direction, and the data are averaged before normalization to account for slow temperature drifts. For the ip and the oopj rotation, the data (open symbols) can be well described by a $\sin^2(\alpha, \beta)$ dependence (solid line), whereas no modulation is visible in the oopt rotation. This angular dependence is characteristic for both the SMR \cite{Chen_Theory_of_SMR,AlthammerSMR} (cf. Eq.\ref{eq:SMR}) and the Hanle magnetoresistance (HMR). \cite{Velez_HanleMR} The HMR is microscopically ascribed to the dephasing of the spin accumulation at the Pt interface, also exists in pure Pt without the adjacent magnetic layer, and scales with the external magnetic field $\mathbf{H}$. In contrast, the SMR depends on the magnetization orientation $\mathbf{m}$, which is expected to be field dependent in a paramagnetic material. However, we do not expect the HMR to contribute significantly to our results, since the magnitude of the HMR for the external magnetic fields used in our measurements ($\mu_0H\leq\SI{2}{\tesla}$) is reported to be $\Delta\rho_{\mathrm{HMR}}/\rho_0\approx2.5\times10^{-6}$ and therefore roughly one order of magnitude smaller than the MR ratio observed here \cite{Velez_HanleMR} $-\Delta\rho/\rho_0=-3\times10^{-5}$. Therewith, $\Delta\rho/\rho_0$ of the noncrystalline YIG/Pt bilayers on YAG is roughly one order of magnitude smaller than in comparable samples featuring a crystalline, ferrimagnetic YIG layer. \cite{AlthammerSMR,Nakayama_Proximity} Similar measurements  performed on reference non-crystalline YIG/Pt bilayers, in particular also on the ones on Si/SiO$_2$ substrates, show the same dependencies, with a magnetoresistance in the same order of magnitude, further supporting the fact that it is indeed the non-crystalline YIG layer that is responsible for the SMR. The angular dependence additionally disputes long-range antiferromagnetic ordering in our bilayers, since a shift of the extrema by $\SI{90}{\deg}$ compared to the SMR introduced in Eq.\ref{eq:SMR} would be expected for an AFM. \cite{Gomonay_AFM_spintronics,Fischer_NiO,Baldrati:2017_NiO, Schlitz_Cr2O3,Hoogeboom_NiO}
	The existence of a magnetoresistance in PMI/Pt heterostructures has already been reported above the Curie temperature for CoCr$_2$O$_4$/Pt bilayers with a MR ratio of  $\Delta\rho/\rho_0<2\times10^{-6}$ by Aqeel \textit{et al.}\cite{Aqueel_SMR_paramagnetic-Phase} and above the Neel temperature for Cr$_2$O$_3$/Pt bilayer structures by Schlitz \textit{et al.} \cite{Schlitz_Cr2O3} with $\Delta\rho/\rho_0>1\times10^{-4}$. It has also been reported that no MR was detectable in paramagnetic Gd$_3$Ga$_5$O$_{12}$ (GGG)/Pt heterostructures at room temperature. In contrast to the other systems, the magnetic moment of GGG has its origin in the $4f$ electrons (vs. $3d$ electrons), which have been suggested not to couple well to the spin accumulation of the Pt layer due to their strong localization.\cite{Schlitz_Cr2O3} Therewith, the reported values of the MR ratio in paramagnetic phases vary between zero and nearly the MR of crystalline YIG/Pt bilayers.\cite{Aqueel_SMR_paramagnetic-Phase,Schlitz_Cr2O3} Our data fall well within this range. We, however, address a real paramagnet and not the paramagnetic phase of a system that orders at lower temperatures.
	
	\begin{figure}
		\includegraphics[width=\linewidth]{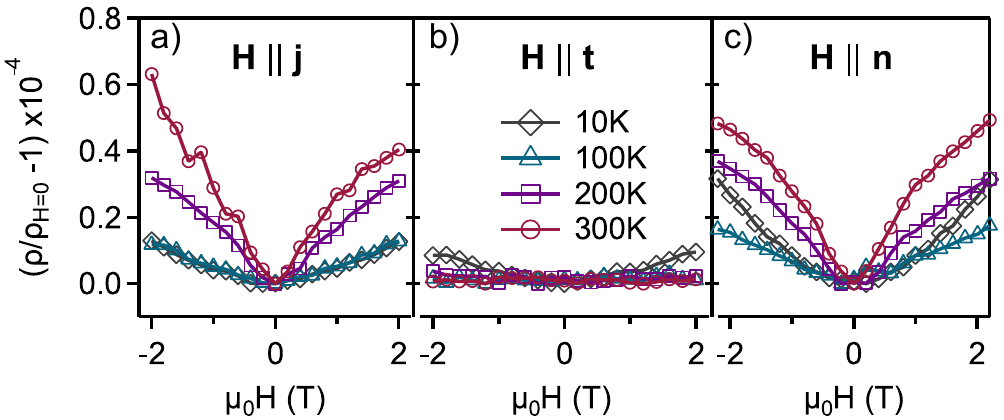}
		\caption{Field-dependent magnetoresistance measurements at different temperatures for \textbf{H}$||$\textbf{j} (panel a), \textbf{H}$||$\textbf{t} (panel b) and \textbf{H}$||$\textbf{n} (panel c). Measurements were performed at different temperatures $\mathrm{T}=10\,$K, $100\,$K, $200\,$K and $300\,$K, which are given by the open gray rhombi, blue triangles, violet squares and red circles, respectively.}
		\label{fig:FDMR}
	\end{figure}
	Field-dependent measurements were performed on the same sample and in the same experimental setup as described before. The resistivity ratio $\rho/\rho_{\mathrm{H}=0}-1$ for $\mathbf{H}||\mathbf{j}$ ($\mathbf{H}||\mathbf{t}$, $\mathbf{H}||\mathbf{n}$) is given in Fig.\ref{fig:FDMR}a (b, c). Note that the results are normalized with respect to the value at zero magnetic field to ensure comparability of the data acquired for different temperatures. An increase in the external magnetic field magnitude along the \textbf{j} and \textbf{n} direction leads to an increase in resistivity. For an external magnetic field applied along \textbf{t} direction, however, no substantial modulation of the resistivity is observed. For these results a significant contribution of the HMR is again excluded since the reported resistivity ratio $\Delta\rho/\rho\approx2.5\times10^{-6}$ at $\SI{2}{\tesla}$ is one order of magnitude lower than the values for the same field magnitude in Fig.\ref{fig:FDMR}.\cite{Velez_HanleMR} For low temperatures, an increase in the external magnetic field does lead to a parabolic increase in the \textbf{t} and \textbf{n} direction that we ascribe to a Kohler's rule type ordinary magnetoresistance in the Pt layer. \cite{Kohler_Widerstand_in_Metallen} Additionally, weak antilocalization has been reported to occur in Pt thin films on various substrates for temperatures below $\SI{50}{\kelvin}$  which might give an additional contribution to the field dependency. \cite{Hoffmann_MRinPtatlowT,Velez_HanleMR,Niimi_WAL_Pt} No saturation of the resistance can be observed in our samples, even for the highest magnetic fields that can be applied in our setup ($\pm\SI{2}{\tesla}$ in the \textbf{j} and \textbf{t} direction and $\pm\SI{6}{\tesla}$ in the \textbf{n} direction) and the lowest temperatures accessible ($\SI{10}{\kelvin}$).
	
	\begin{figure}
		\includegraphics[width=\linewidth]{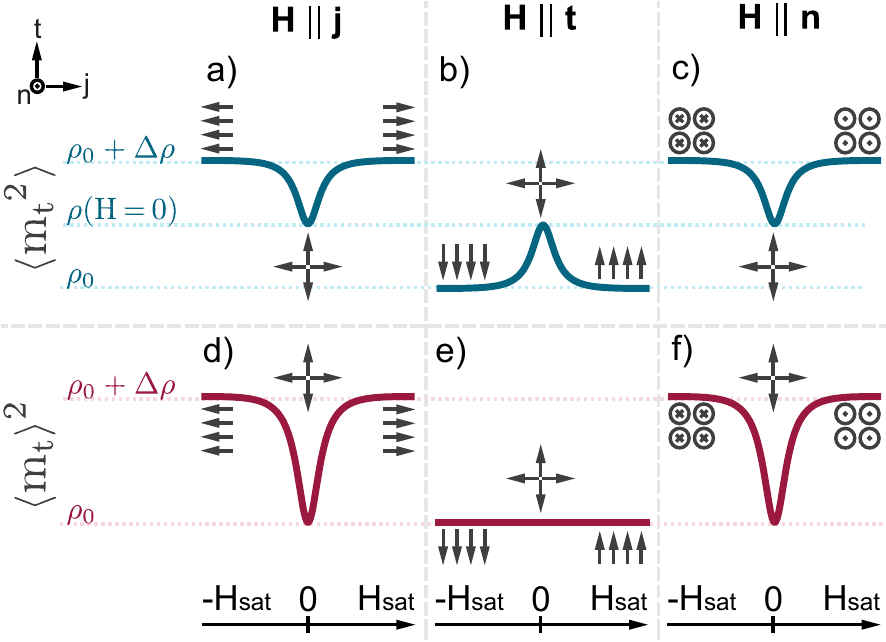}
		\caption{Assuming that all magnetic moments in a paramagnet contribute individually to the SMR (i.e. $\langle\mathrm{m_t}^2\rangle$), one obtains the evolution of the resistivity sketched as a blue line for $\mathbf{H}||\mathbf{j}$ (a), $\mathbf{H}||\mathbf{t}$ (b), $\mathbf{H}||\mathbf{n}$ (c). Presuming that the SMR is dependent on the net magnetization (i.e. $\langle\mathrm{m_t}\rangle^2$), the expected magnetoresistive response is shown as a red line in panel d, e and f, for $\mathbf{H}||\mathbf{j}$, $\mathbf{H}||\mathbf{t}$ and $\mathbf{H}||\mathbf{n}$, respectively. The external magnetic field applied in each direction is indicated by the arrows below the panels. The alignment of the magnetic moments for different external field amplitudes is given by the arrows on above and below the resistance curves.}
		\label{fig:BallOfMoments}
	\end{figure}
	We now compare the SMR expected from a model that considers an ensemble of non-interacting moments (Fig.\ref{fig:BallOfMoments}a-c) with the SMR arising in a model that addresses the total net moment (Fig.\ref{fig:BallOfMoments}d-f). To that end, we consider external magnetic fields applied along \textbf{j}, \textbf{t} and \textbf{n} direction as schematically shown above the panels in Fig.\ref{fig:BallOfMoments}. In a paramagnet, the moments at zero magnetic field are unordered and point in random directions, as schematically sketched in Fig.\ref{fig:BallOfMoments}. For a sufficiently large magnetic field $\mathrm{H}\rightarrow \pm\mathrm{H_{sat}}$, though, the majority of the magnetic moments are aligned collinear to the external field. Since the SMR is sensitive to $\mathrm{m_t}^2$, one would expect the smallest resistivity (i.e. $\rho_0$) when all moments are parallel to the \textbf{t} direction and the largest resistivity (i.e. $\rho_0+\Delta\rho$) when all moments are perpendicular to the \textbf{t} direction (c.f. Eq.\ref{eq:SMR}). Thus, $\rho_0$ is the value expected for open boundary conditions (i.e. no magnetic layer), while the introduction of a magnetic layer to the system can only lead to an increase in the resistivity depending on its magnetization orientation.\cite{Chen_Theory_of_SMR}
	
	Presuming non-interacting moments in the PMI, the contribution of each moment to the SMR is considered separately. Thus, $\mathrm{m_t}^2$ in Eq.\ref{eq:SMR} has to be understood as $\langle\mathrm{m_t}^2\rangle$. Applying sufficiently large magnetic fields leads to a saturation of the resistivity at a minimum value $\rho_0$ for $\mathbf{H}||\mathbf{t}$ (Fig.\ref{fig:BallOfMoments}b) and maximum value $\rho_0+\Delta\rho$ for $\mathbf{H}||\mathbf{j},\mathbf{n}$ (Fig.\ref{fig:BallOfMoments}a, c). In turn, this implies that for zero magnetic field an intermediate resistance $\mathrm{\rho}({\mathrm{H}=0})$ is expected with  $\mathrm{\rho}_0<\mathrm{\rho}(\mathrm{H}=0)<\mathrm{\rho}_0+\Delta\mathrm{\rho}$, since some but not all magnetic moments are collinear to the \textbf{t} direction (panel a, b and c of Fig.\ref{fig:BallOfMoments} at $\mathrm{H}=0$).
	
	Assuming that the SMR depends on the total net magnetization in the paramagnetic layer instead, $\mathrm{m_t}^2$ in Eq.\ref{eq:SMR} has to be understood as $\langle\mathrm{m_t}\rangle^2$. Hence, no magnetic moment is expected for zero external magnetic field, which leads to a vanishing SMR and therefore to the minimum resistivity value $\rho_0$ (Fig.\ref{fig:BallOfMoments}d-f). Applying sufficiently large external magnetic fields along $\mathbf{j}$ and $\mathbf{n}$ leads to an increase in resistivity up to the saturation value of $\rho_0+\Delta\rho$ as is shown in Fig.\ref{fig:BallOfMoments}d and e. In contrast, no change in the resistivity is expected for fields applied along $\mathbf{t}$ direction (Fig.\ref{fig:BallOfMoments}e).
	
	Comparing the expected behavior from Fig.\ref{fig:BallOfMoments} with the field-dependent measurements in Fig.\ref{fig:FDMR}, shows that the measurements do not agree with the SMR stemming from an ensemble of non-interacting moments (see Fig.\ref{fig:BallOfMoments}a-c). Instead, the measurements corroborate that the SMR in the paramagnetic YIG/Pt bilayers is determined by the total net magnetization of the system (cf. Fig.\ref{fig:BallOfMoments}d-e). This result contradicts previous findings in compensated garnets and antiferromagnets where the results were explained by taking into consideration the magnetizations of the different sublattices separately. \cite{Dong_GdIG,Ganzhorn_GdIG,Fischer_NiO,Hoogeboom_NiO} To confirm the saturation of the effect and to corroborate our observations, further experimental work on different PMI materials, as well as at higher external magnetic fields and/or lower temperatures, is necessary.
	
	In summary, we have studied the magnetoresistive response in noncrystalline, paramagnetic YIG/Pt heterostructures. Upon rotating the external magnetic field at a fixed magnitude in different planes, we observe a magnetoresistance with the characteristics of the spin Hall magnetoresistance with a magnitude of $\lvert\Delta\rho/\rho\rvert=3\times 10^{-5}$ at $\mu_0\mathrm{H}=\SI{2}{\tesla}$ and $\SI{200}{\kelvin}$. Field-dependent measurements show an increase of resistivity for an increasing magnetic field along \textbf{j} and \textbf{n} direction, whereas no change of the resistance is observed for fields applied along \textbf{t} direction. No saturation is detected for the maximum magnetic fields accessible in our setup. Furthermore, we propose two possible models for the origin of the SMR in a simple paramagnetic insulator/Pt heterostructure taking into consideration either an ensemble of non-interacting moments (i.e. $\langle\mathrm{m_t}^2\rangle$), or the total net magnetization (i.e. $\langle\mathrm{m_t}\rangle^2$). Comparing the experimentally observed signature with those models, we find that the data are better described in terms of the total moment picture. Thus, we conclude that in a paramagnetic insulator, the moments do not contribute individually, but in a collective net fashion to the SMR.
	
	We thank K.\ Nenkov and B.\  Weise for technical support. We acknowledge financial support by the Deutsche Forschungsgemeinschaft via SPP 1538 (project GO 944/4).
	
	\bibliography{references.bib}
	
\end{document}